%% file: main.tex
\documentclass[sigconf,nonacm,authorversion]{acmart}
\AtBeginDocument{%
  }

\copyrightyear{2025}
\acmYear{2025}
\setcopyright{rightsretained}
\acmConference[CHI EA '25]{Extended Abstracts of the CHI Conference on Human Factors in Computing Systems}{April 26-May 1, 2025}{Yokohama, Japan}
\acmBooktitle{Extended Abstracts of the CHI Conference on Human Factors in Computing Systems (CHI EA '25), April 26-May 1, 2025, Yokohama, Japan}\acmDOI{10.1145/3706599.3719757}
\acmISBN{979-8-4007-1395-8/2025/04}

\begin{document}

\title{Vipera: Towards systematic auditing of generative text-to-image models at scale}

\author{Yanwei Huang}
\email{yanweih@andrew.cmu.edu}
\orcid{0009-0001-9453-7815}
\affiliation{%
  \institution{Human-Computer Interaction Institute\\Carnegie Mellon University}
  \city{Pittsburgh}
  \state{Pennsylvania}
  \country{USA}
}

\author{Wesley Hanwen Deng}
\email{hanwend@cs.cmu.edu}
\orcid{0000-0003-3375-5285}
\affiliation{%
  \institution{Human-Computer Interaction Institute\\Carnegie Mellon University}
  \city{Pittsburgh}
  \state{Pennsylvania}
  \country{USA}
}

\author{Sijia Xiao}
\email{xiaosijia@cmu.edu}
\orcid{0009-0004-5889-1376}
\affiliation{%
  \institution{Human-Computer Interaction Institute\\Carnegie Mellon University}
  \city{Pittsburgh}
  \state{Pennsylvania}
  \country{USA}
}

\author{Motahhare Eslami}
\email{meslami@andrew.cmu.edu}
\orcid{0000-0002-1499-3045}
\affiliation{%
  \institution{Human-Computer Interaction Institute\\Carnegie Mellon University}
  \city{Pittsburgh}
  \state{Pennsylvania}
  \country{USA}
}

\author{Jason I. Hong}
\email{jasonh@cs.cmu.edu}
\orcid{0000-0002-9856-9654}
\affiliation{%
  \institution{Human-Computer Interaction Institute\\Carnegie Mellon University}
  \city{Pittsburgh}
  \state{Pennsylvania}
  \country{USA}
}

\author{Adam Perer}
\email{adamperer@cmu.edu}
\orcid{0000-0002-8369-3847}
\affiliation{%
  \institution{Human-Computer Interaction Institute\\Carnegie Mellon University}
  \city{Pittsburgh}
  \state{Pennsylvania}
  \country{USA}
}


\begin{abstract}
  Generative text-to-image (T2I) models are known for their risks related such as bias, offense, and misinformation. Current AI auditing methods face challenges in scalability and thoroughness, and it is even more challenging to enable auditors to explore the auditing space in a structural and effective way. Vipera employs multiple visual cues including a scene graph to facilitate image collection sensemaking and inspire auditors to explore and hierarchically organize the auditing criteria. Additionally, it leverages LLM-powered suggestions to facilitate exploration of  unexplored auditing directions. An observational user study demonstrates Vipera's effectiveness in helping auditors organize their analyses while engaging with diverse criteria. 
\end{abstract}





\begin{teaserfigure}
  \centering
  \includegraphics[width=0.81\linewidth]{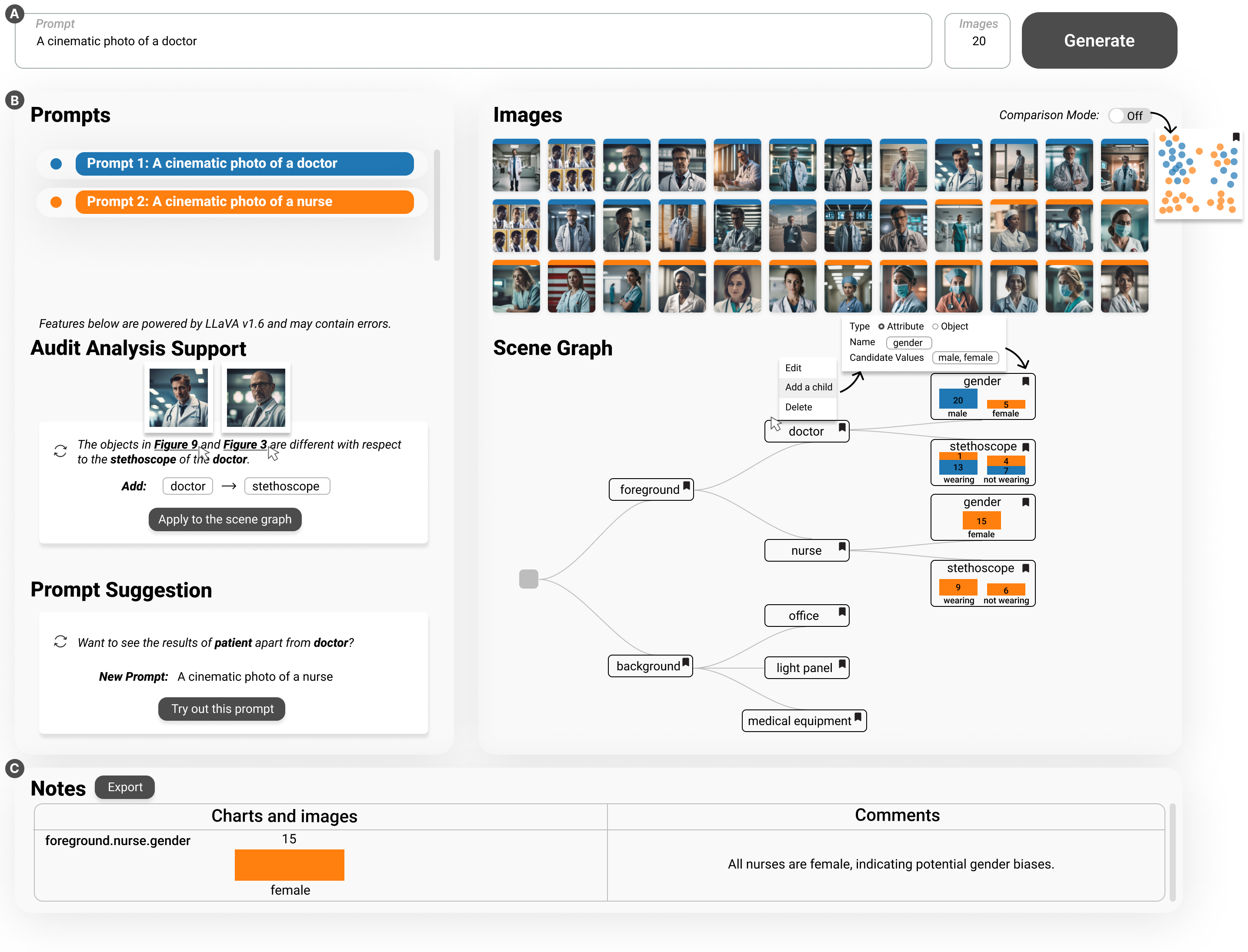}
  \caption{The Vipera interface. (A) The input box for creating prompts and specifying the number of images. (B) The analysis view showing the prompts, images, scene graph, and auditing suggestions. (C) The note view for composing an auditing report. }
  \Description{The interface displays a tool for generating images based on prompts. On the left side, a text box shows the prompt "A cinematic photo of a doctor," with options for audit analysis support and prompt suggestions. The central area features a collection of images related to the prompt, depicting various scenes with doctors and medical settings. To the right, a scene graph outlines the relationships between elements, including "doctor," "nurse," "office," and "medical equipment." Notes at the bottom indicate details about gender representation among nurses.}
  \label{fig:system}
\end{teaserfigure}


\maketitle

\input{sections/01_introduction}
\input{sections/02_related_work}
\input{sections/03_design_goals}

\input{sections/04_vipera}
\input{sections/05_evaluation}

\input{sections/06_conclusion}

\begin{acks}
The research was supported by National Science Foundation (NSF) program on Fairness in AI in collaboration with Amazon under Award No. IIS-2040942, as well as an award from Notre Dame–IBM Technology Ethics Lab. We would like to thank all the anonymous reviewers for their constructive comments.
\end{acks}

\bibliographystyle{ACM-Reference-Format}
\bibliography{main}










\end{document}

%% file: sections/01_introduction.tex
\section{Introduction}
Generative text-to-image (T2I) models are gaining popularity for their ability to enhance the efficiency and expressiveness of creative activities. However, the associated risks are significant; these models can produce images that may perpetuate biases, cause offense, or disseminate misleading information~\cite{t2i-bias, qi2023visual}. This underscores the process of detecting and addressing these issues, or \textit{AI auditing}, where auditors analyze system outputs for adversarial inputs~\cite{sandvig2014auditing, metaxa2021auditing, birhane2024ai}.

Today, auditing generative T2I models in a systematic and scaled way remains challenging.
Manual evaluation of these systems can be time-consuming, often limiting assessments to a small number of prompts or outputs, which can hardly uncover the comprehensive issues of the system. In response, recent research has explored tools and processes to engage end users in AI audits~\cite{enduseraudits, everydayaudit, cabrera2021discovering, deng2023understanding, maldaner2024mirage, claire2024designing}. Nevertheless, developers face challenges in exploring potentially productive auditing directions and gathering scaled data for each identified issue~\cite{enduseraudits, deng2025weaudit}. Reviewing and aggregating audit reports also remains a demanding task \cite{cabrera2021discovering, deng2023understanding, ojewale2024towards}

To tackle these challenges, automated approaches have been proposed, often involving AI-supported output labeling and applied to textual models~\cite{who-validates, Rastogi2023LLM-audit-LLM, chainforge}. However, two primary obstacles persist when it comes to auditing T2I models. First, images encompass a broad range of semantics, resulting in a vast auditing space that is difficult to characterize with a concise list of criteria as often used for evaluating texts\cite{sangho2024luminate}. For example, an image may feature multiple characters, each of whom can be evaluated based on the clothing, and the clothing can be further assessed by its cultural nature and relationship to the character's profession. To systematically explore this space, it is crucial to keep auditors aware of the \textit{unknown unknowns} - criteria that are yet to be explored \cite{kiela2021dynabench, deng2025weaudit}. Additionally, our observations from formative study with five auditors indicate that many auditors often rely on intuition or personal experience during the auditing process without a formal approach. There is a lack of structured methodologies for navigating and exploring this extensive auditing space.

In this paper, we propose Vipera\footnote{Vipera stands for "\textbf{V}isual \textbf{I}ntelligence-\textbf{P}romoted \textbf{E}nd Use\textbf{r} \textbf{A}uditing"}, an interactive system for streamlining and enhancing the systematicness of large-scale T2I model auditing. Vipera facilitates structured multi-faceted analysis through the visual cue of \textit{scene graph}, where the auditing criteria are organized hierarchically and associated with scenic semantics of images. Additionally, Vipera incorporates LLM-powered auditing suggestions to uncover new avenues for analysis and facilitate exploration within the auditing space.
Through a pilot user study with five T2I auditors, we demonstrated initial evidences on the Vipera's usability and effectiveness in helping auditors structurally organize their auditing, while exploring diverse auditing criteria and prompts through drawing inspirations LLM-powered auditing suggestions. We hope that Vipera could reveal the abundant opportunities and contribute innovative insights to the emerging field of GenAI auditing and human-AI collaborative auditing.

%% file: sections/02_related_work.tex
\section{Related Work}
\subsection{Auditing generative AI}
Generative AI (GenAI) systems have sparked increasing societal concerns due to potential harmful behaviors, such as social biases and violence~\cite{algorithm-harms}. This has led to increased focus on detecting and mitigating these issues through benchmarks like SafetyBench~\cite{zhang2023safetybench} and AgentHarm~\cite{andriushchenko2024agentharm}. However, the limited diversity of inputs and contexts in these benchmarks hampers real-world performance assessments. Recent research emphasizes the value of conducting AI audits to ensure more safe and responsible GenAI, often even involving general end users in the auditing process, as they can uncover overlooked cases and provide insights~\cite{everydayaudit,enduseraudits,crowdsourced-failure-reports,deng2025weaudit}. Despite this potential, prior research in HCI show that crowdsourced auditing remains challenging and costly due to data size and limited criteria \cite{cabrera2021discovering, deng2023understanding}. Recent studies have begun to focus on scalable auditing of generative AI, primarily addressing two key challenges. First, evaluating numerous outputs within a limited timeframe can be difficult. To tackle this, human-in-the-loop auditing has emerged, utilizing large language models for automatic evaluations while organizing results for user interpretation~\cite{llm-as-a-judge, who-validates, llm-audit-llm}. Additionally, annotations and visualizations have been introduced for better understanding of the auditing results~\cite{gero-atscale, adversaflow}. Second, early auditing tools often relied on inflexible quantitative metrics, but recent efforts allow users to define their own criteria~\cite{evallm}.

Furthermore, aforementioned work has mainly focused on auditing text-to-text models; extending them to text-to-image models presents additional challenges. Images contain rich semantic information (e.g., objects and styles) that can lead to a diverse range of auditing criteria, yet there is a notable lack of tools to assist auditors in interpreting auditing results for iterative assessments in the image context \cite{pettersson1993visual}. To address this, Vipera integrates visual cues and AI-driven auditing suggestions, facilitating a structured and systematic exploration of auditing criteria. 

\subsection{Visual analytics for image collection sensemaking}
Sensemaking of large image collections plays a key role in auditing T2I models at scale, and visual analytics approaches have proven effective in this context~\cite{afzal2023visualization}. One common method involves clustering images based on pixel or vector embeddings for visualization, which supports various analytical tasks such as search and exploration, comparison, and visual summarization~\cite{ii20,schmidt2013vaico,pan2019content}. To incorporate image semantics alongside visual features, recent works identify semantic objects within images and extract concise textual representations—such as keywords, descriptions, or captions. These text analysis results facilitate tasks like semantic categorization and pattern mining~\cite{xie2018semantic,yee2003faceted,li2024visual}. However, characterizing the relationships between objects using plain text can be challenging. To address this, scene graphs have emerged as a more expressive representation, where nodes denote semantic objects and edges illustrate their relationships. This representation enhances detailed understanding of image collections and improves analysis performance~\cite{fang2017narrative,johnson2015image}.

With the advent of generative T2I models, a recently emerged task is to visualize the relationship of between prompts and corresponding images. For instance, DreamSheets organizes prompts and images in a spreadsheet layout to facilitate comparison \cite{dreamsheets}. Similarly, PrompTHis introduces an \textit{image variant graph} to illustrate the semantic transitions of image clusters as users create new prompts \cite{guo2024prompthis}. However, these approaches often struggle to support sensemaking in large-scale image sets, where variations among outputs for a single prompt can lead to overly condensed visual clusters. In response, Vipera embeds bar charts within a tree-based scene graph, revealing distributional patterns for images related to individual prompts and allowing for comparisons across prompts. Additionally, it leverages the scene graph as a visual aid to foster systematic and structured AI auditing while providing inspiration to auditors. 

%% file: sections/03_design_goals.tex
\section{Design Study \& Goals}

\subsection{Formative study}
To inform Vipera's design, we conducted a formative study with two main objectives: (1) understanding the common practices and challenges auditors face when evaluating generative text-to-image (T2I) models, and (2) testing our hypothesis that a scene graph as a visual aid could enhance insights and structure in the auditing process.

\subsubsection{Study setup} \quad

\textbf{Participants.} The study involved five participants (P1-5) recruited from our collaborating institutions, including three Ph.D. students, one Postdoctoral researcher, and one software engineer, all of whom had prior experience in auditing generative AI or algorithms. 

\begin{figure*}[t]
  \centering
  \includegraphics[width=0.8\linewidth]{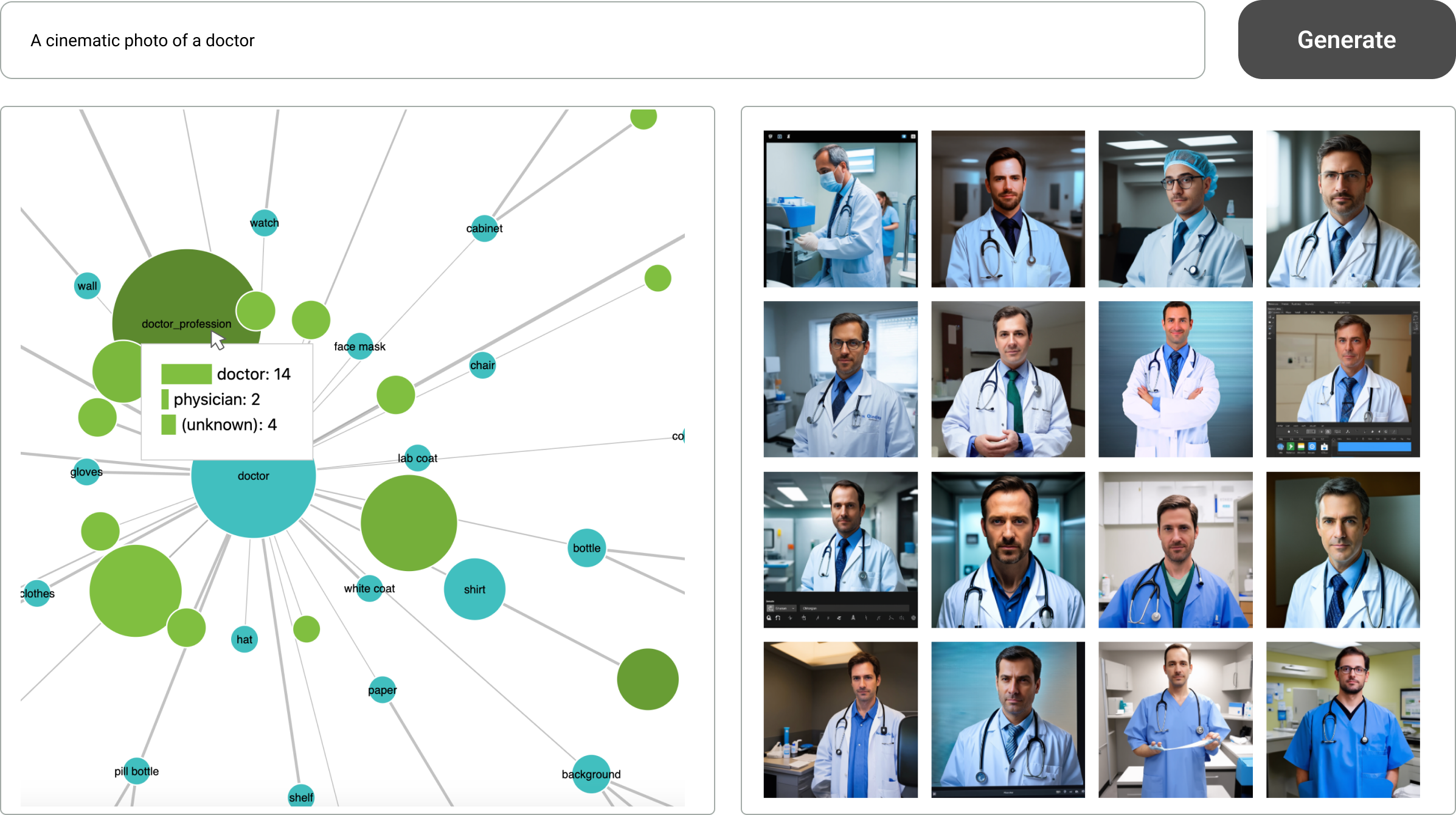}
  \caption{The ViperaBase prototype used in the formative study, showing the generated images (right) and a scene graph (left) for the user prompt (top). The scene graph is a node-link diagram where nodes indicate objects (or their attributes) within the images and  edges indicate the semantic relationships. Bar charts will be shown when hovering on attribute nodes. }
  \Description{The image displays a visual representation of terms related to "a cinematic photo of a doctor." On the left, a network graph shows the frequency of related terms, with "doctor" being the most prominent. On the right, a collage features various images of male and female doctors in clinical settings, each wearing medical attire and stethoscopes. The layout highlights the connection between the terms and the visual depictions of healthcare professionals.}
  \label{fig:formative}
\end{figure*}

\textbf{Prototypes.}
Following parellel prototyping \cite{dow2010parallel}, we designed two prototypes for the formative study, referred to as ViperaBase and Baseline in the following discussions. As illustrated in \autoref{fig:formative}, ViperaBase features a scene graph view, i.e., the zoomable node-link diagram, alongside a prompt input box and image view,  whereas the Baseline includes all views except the scene graph view. In this view, blue nodes represent objects, and green nodes indicate their attributes, with labeled edges illustrating relationships. The size of each node reflects the number of images containing that object or attribute. Hovering over an attribute node displays a bar chart of evaluation results. Images were generated by Stable Diffusion v3-medium, while the scene graph was created using LLaVA v1.5.

\textbf{Protocol.} We first conducted semi-structured interview to explore participants' current auditing practices. Participants were then tasked with auditing the Stable Diffusion model using both prototypes in order (Baseline first, followed by ViperaBase). Each auditing session lasted five minutes, during which participants selected a prompt from a pre-prepared list to generate 30 images, explored these images for as many insights as possible, and noting their insights and evidence in brief phrases or sentences. After interacting with each prototype, participants shared their reasoning processes. Finally, they provided feedback on their perceptions of both prototypes and requirements for Vipera. All study sessions were conducted online, and the average duration of each session was approximately 50 minutes. The study is apporved by IBR.

\subsubsection{Findings} Our primary observations and insights from the study are summarized as follows.

\textbf{Typical scales and motivations for T2I auditing. } All participants reported that the scale of auditing varies based on the specific task. While the most common average scale for auditing was reviewing around 10 images at once, some participants indicated that they had auditing hundreds of images simultaneously in certain cases. Motivations for large-scale auditing differed among participants: auditors who were also developers often engaged in systematic testing. For example, P1 reported generating 30 images for 10 different categories or topics. In contrast, non-developer auditors preferred smaller scales typical of their daily work but would analyze a larger image corpus if the initial images have issues or did not meet their requirements.

\textbf{Auditing practices.} Generally, the most common method for auditing was described as ``just looking at the images''. However, participants acknowledged the time-consuming nature of examining images at scale. Specifically, P1 and P4 utilized generative AI tools with visual understanding capabilities, such as GPT-4 and LLaVA~\cite{llava}, to gain observations or insights. Nonetheless, these tools presented challenges, including inaccuracies or lackluster results (as noted by P4) and visual anomalies—like distorted human hands—that could only be identified by human reviewers (observed by P1).

\textbf{Auditing criteria.} The study shows that ViperaBase helped users identify unnoticed abnormalities compared to Baseline. Participants explained that this was because the additional scene graph serves as a ``check-list'', enabling them to examine the images in a disciplined manner, compared to merely relying on their experience and intuition in their usual workflow. Moreover, all of them agreed that the node size encoding was helpful in uncovering minority objects, while comparing the sizes between adjacent nodes was also likely to yield interesting insights. However, three participants mentioned that the dense scene graph was too time-consuming to go through, and they hoped to customize the graph, such as adding evaluation criteria related to the keywords in their prompts.


\subsection{Design goals}
Based on the findings and feedback from the design study, we have derived the following design goals (DGs):

\begin{itemize}
    \item DG1: Leverage interactive visual aids for structured, customizable auditing and effective result sensemaking.  

    \item DG2: Enable intuitive comparison of how the distribution of large scale image outputs changes across prompts.

    \item DG3: Promote divergent thinking by highlighting image details that auditors might have overlooked.

    \item DG4: Inspire auditors to explore prompts that could lead to insightful audit outcomes.

    \item DG5: Assist auditors in documenting, synthesizing, communicating, and acting upon their audit findings.
\end{itemize}

%% file: sections/04_vipera.tex
\section{The Vipera System}
This section presents the design and implementation details of Vipera. As shown in ~\autoref{fig:system}, three major components are included: an input box (A) for creating prompts, an analysis view (B) for customizing the auditing process and analyzing the results, and a note view (C) for report generation. 

\subsection{Analysis view}
The analysis view consists of four subviews: prompts, images, scene graph, and audit suggestions.

\textbf{Prompts.} The prompts view displays all prompts created by the user, each encoded by a different color. By default, the results of all prompts will be displayed and analyzed. Users are allowed to compare specific prompts by deselecting unwanted prompts through clicks (\textbf{DG2}).

\textbf{Images.} The image view shows the thumbnails of the generated images, with the border color indicating the corresponding prompt. Users can click on the images to see the full-sized version as well as the labels for the auditing criteria defined in the scene graph. To further facilitate comparing the distribution of images for different prompts (\textbf{DG2}), users are allowed to switch to the \textit{comparison mode}, where images are clustered based on their labels and visualized in a scatterplot. They may also click on individual nodes to navigate to the corresponding images.

\textbf{Scene graph.} The scene graph serves as both a visual summary of the semantic contents within the images and a visual cue for users to organize their auditing (\textbf{DG1}). Objects in the scene graph are represented as nodes and organized in a tree layout based on their semantic relationship. To add an auditing criterion, users can right click on a node and select to add an attribute node to the child of the current node in the result modal. The added attribute will then be used to evaluate and label the images. We further allow users to set predefined candidate label values (e.g., \textit{male} and \textit{female} for \textit{gender}) to specify wanted categories for classification and prevent the generative AI from generating a too diverse range of models. The result labels will be visualized in stacked bar charts with colors indicating the corresponding prompts. In addition, users are also allowed to add regular object nodes to the scene graph, to which attribute nodes can be further added for evaluation.

\textbf{Suggestions.} Informed by \textbf{DG3} and \textbf{DG4}, Vipera provides two classes of auditing guidance. The \textit{audit analysis support} highlights two images with notable differences and suggests auditing additional auditing criteria as nodes. Meanwhile, the \textit{prompt suggestion} promotes divergence thinking by suggesting insightful prompts, replacing words or phrases in existing prompts. Upon users deciding to try out the prompt, Vipera will duplicate the existing auditing criteria while generating new images, allowing for an effective comparison of results between different prompts.

\subsection{User workflow and technical pipeline}
\begin{figure}[t]
  \centering
  \includegraphics[width=\linewidth]{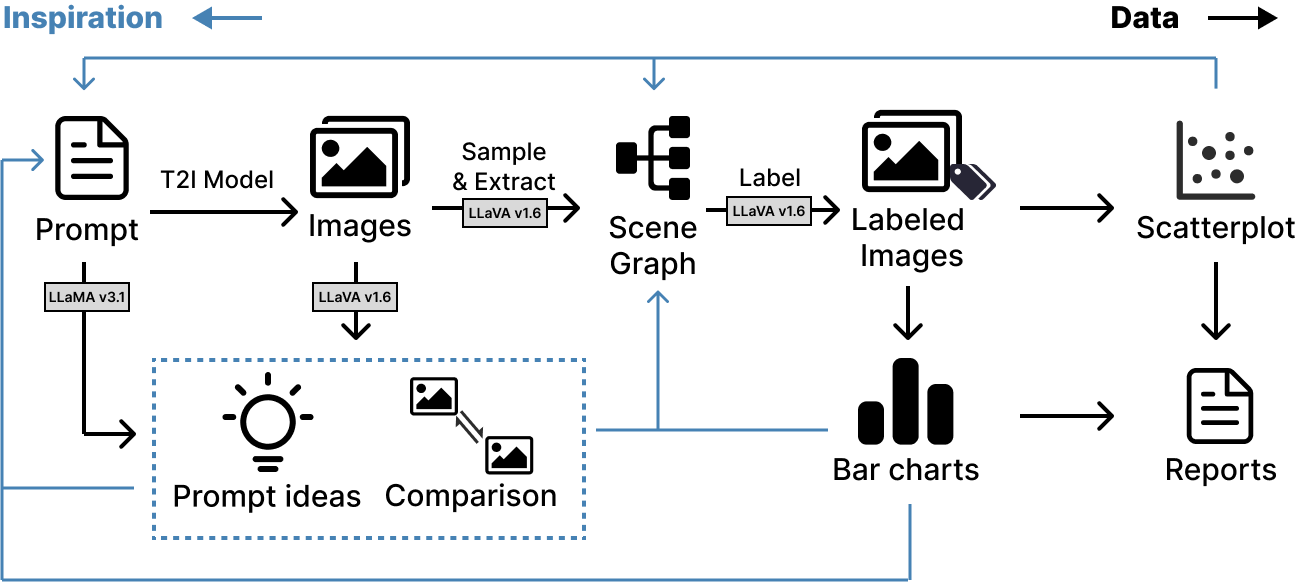}
  \caption{Vipera's technical pipeline. Black edges indicate the data flow and blue ones indicate the inspiration flow for iterative auditing. }
  \Description{The flowchart illustrates a process for generating and analyzing images using the T2I model. It begins with a prompt that leads to the creation of images. These images are then sampled and extracted to form a scene graph. The scene graph is labeled to produce labeled images, which can be visualized in a scatterplot or represented in bar charts. Additionally, there is a section for prompt ideas and comparisons, indicating a feedback loop for inspiration and refinement of prompts. The entire process is driven by data, which flows from right to left in the diagram.}
  \label{fig:workflow}
\end{figure}

Vipera has been developed and deployed as a web-based application. ~\autoref{fig:workflow} shows the workflow of Vipera\footnote{All prompts used in the pipeline are provided in the appendix.}.  After generating images based on the user's prompt and the T2I model being audited, Vipera utilizes LLaVA v1.6 to extract tree-like scene graphs from a sampled set of the image collection and aggregates these graphs into a single comprehensive representation. Note that we use a tree layout with a maximum of 5 leaf nodes to avoid overwhelming the users (\textbf{DG1}). With the scene graph serving as a visual summary, users can add object or attribute nodes (i.e., auditing criteria) to this visual cue. These additions are then used to label the images through LLaVA v1.6. Subsequently, Vipera generates two types of visualizations to aid in result sensemaking and stimulate further auditing: a scatterplot that clusters the images based on the assigned labels, and bar charts displaying the distribution of labels for each attribute node. With these visualizations as inspirations, users may iteratively modify the scene graph or try out new prompts for detailed auditing, illustrated as blue arrows in ~\autoref{fig:workflow}. Additionally, they may adopt the suggestions from Vipera at any time during the auditing. Specifically, the prompts suggestions are generated by LLaMA v3.1, while the criteria suggestions are from LLaVA v1.6 after feeding it with several randomly selected image pairs. 
Furthermore, users may bookmark any visualizations of interest or images during their work, which can be later exported as an auditing report (\textbf{DG5}).

%% file: sections/05_evaluation.tex
\section{Observational Study}
We conducted an observational user study to evaluate the usability and effectiveness of Vipera.

\subsection{Study setup}
\textbf{Participants.} We recruited five participants (P1-P5) from our institution. All participants had rich experience of using generative T2I models and three of them had engaged auditing them either voluntarily or in a paid manner. Two of them were also involved in the formative study and had used an earlier version of Vipera.

\textbf{Procedure.} After introducing the study background and objectives, we asked the participants to sign the consent form. Participants were then provided a 5-minute tutorial of Vipera's features, and they were given another 5 minutes to get familiar with the system. Afterwards, participants were provided with 30 minutes to audit the Stable Diffusion XL model with Vipera, during which they were asked to get as many auditing insights as possible and note down necessary evidence for their report using Vipera's bookmark features. There were no restrictions on the prompts or auditing criteria they might create. The study concluded with a 15-minute interview. The total duration of the study was approximately 60 minutes and each participant was compensated \$10.

\subsection{Results}
In general, participants reached a consensus in Vipera's usability and they recognized the intuitiveness and effectiveness of Vipera's visualizations in facilitating image sensemaking. Next, we summarize our major observations.

\textbf{Vipera enhanced both the exploration of various auditing criteria and the validation of user hypotheses.} Our observations indicated that participants commonly began by assessing well-known social biases, such as gender and race. Over time, they shifted their focus to the diversity of images, experimenting with a wider range of criteria after noticing unexpected patterns during the audit analysis or while inspecting images, such as the presence of a man's beard or clothing. Notably, only a few participants examined features unrelated to human characters. Two participants attributed this to time constraints, while another two expressed skepticism regarding the significance or relevance of these features. Additionally, we noted that P2 adopted a unique workflow, establishing clear auditing goals and brainstorming extensively before beginning the process. For example, he initially suspected that the model might misinterpret the phrase ``mixed-double matches'' in the prompt, potentially leading to an incorrect count of players in the images. He thus formulated a criterion based on player number, utilizing Vipera as a tool to validate his hypotheses.

\textbf{Vipera facilitated prompt comparison though the prompts were not always comparable.} All participants agreed that it was easy for them to compare the label distribution of images for selected prompts in Vipera, which was mostly attributed to the stacked bar charts in the scene graph. We surprisingly observed that most participants seldom checked the scatterplot in the comparison mode. According to the interview, one major reason was that the scene graph was believed to be useful enough, playing a dominant role in participants' workflows. Meanwhile, P3 explained that he was not fully convinced by the visual clusters in the scatterplot where some criteria that were not directly comparable should not be considered. For instance, it was natural that \textit{doctors} and \textit{patients} would have significantly different \textit{motions}, which could mean nothing and hardly contribute to insights. He therefore suggested an customizable and interactive mode of comparison where the prompts, images, and auditing criteria could be specified by users. 

\textbf{Vipera improved the breadth of auditing, yet user-directed depth-first auditing support was desired. } Participants held divided opinions when asked about their perceptions of the auditing process. Most of them believed that Vipera made their auditing more \textit{ordered} and \textit{comprehensive} due to the inspirations from the scene graph and the auditing suggestions. However, three participants expressed their inconfidence throughout the process. P1 and P3 were especially worried about the \textit{endlessness} of auditing, arguing that the auditing space was still too large despite Vipera's features. Furthermore, while acknowledging the inspirations drawn from Vipera, P1 was disappointed that the Vipera's auditing suggestions  mostly focused on the \textit{breadth} of auditing rather than \textit{depth}, i.e. diving deeper into a specific direction. Similarly, P5 hoped that the system could base the suggestions on his previous interactions, such as recommending criteria relevant to the inspected label when hovering the mouse on a bar in the bar chart.

\subsection{Discussion}
In this section, we discuss the lessons learned from the design process and participants' feedback, as well as the directions we plan for future work.

\textbf{Record auditing provenance for recalling and audit storytelling.} One participant, P1, suggested that Vipera should not only display the history of prompts but also record ``how a prompt is developed''. He emphasized the importance of maintaining awareness of auditing provenance, as learning from the origins of previous ideas could inspire future auditing efforts. Additionally, P1 and P3 expressed a desire to incorporate this provenance information as well as relevant visual cues into the auditing report to create a more engaging narrative. We plan to implement these features and encourage users to share their auditing reports to foster inspiration among peers~\cite{deng2025weaudit}.

\textbf{Enable fine-grained customization of auditing scope.} The current Vipera automatically tested all criteria for all prompts and images, which might lead to uninsightful findings and unnecessary computing. We plan to address this issue by improving the fine-grained control of interactions, including allowing users to set the application scope for each criterion and compare prompts in specified criteria. Moreover, as users might create significantly distinct prompts, we plan to enable users to create multiple workspaces to categorize their auditing directions or classify the prompts into groups for more focused exploration~\cite{suh2023sensecape}. We also plan to explore ways to better integrate Vipera into auditors’ existing workflows to support a more customized and agile auditing process~\cite{wang2024farsight, deng2022exploring}

\textbf{Improving intent communication in human-AI collaborative auditing.} While most participants were satisfied that the generative AI plays a helpful role in their auditing, their collaboration could be improved in many ways. First, although the current suggestions from generative AI assist participants in determining their next steps, they often lack a broader context. For instance, could generative AI develop a mental model of the potential pitfalls associated with the audited model and outline a comprehensive auditing plan for users? Additionally, can it alert users when they should delve deeper into a specific area or when to adopt a more divergent approach to audit more broadly? We envision creating theories, frameworks, or methodologies through better studying the user needs.
Second, while the accuracy of generative AI in labeling is generally satisfactory, we noticed that some participants tended to accept labeling results without further examination. This reliance may lead to potentially misleading auditing insights. Drawing from recent work ~\cite{who-validates}, we aim to develop strategies that keep users aware of these risks, such as incorporating multiple AI judges as references.

%% file: sections/06_conclusion.tex
\section{Conclusion}
We introduce Vipera, an innovative system designed to enhance the systematic auditing of these models by providing structured multi-faceted analysis and LLM-powered suggestions. Vipera further leverages multiple visual aids including scene graphs, stacked bar charts, and scatterplots, to facilitate intuitive sensemaking of auditing results and assist auditors in streamlining and organizing their auditing. Our observational user study confirms Vipera's usability and effectiveness, demonstrating its potential to help auditors navigate the complex auditing landscape while uncovering new insights. We believe Vipera will contribute significantly to the advancement of end-user auditing and foster more responsible human-AI collaboration in creative applications.